\UseRawInputEncoding
\documentclass[preprint,prc,amsmath,amssymb,superscriptaddress,noffotinbib,preprintnumbers]{revtex4-1}
\usepackage{url,color,colordvi,epsfig,pstricks}
\usepackage{hyperref}
\usepackage{cleveref}
\usepackage{graphicx,bm,dcolumn}
\hypersetup{
    colorlinks=true, 
    linktoc=all, 
    linkcolor=blue, 
    citecolor=blue}

\begin{document}

\title{Potential Constraints to Neutrino - Nucleus Interactions Based on Electron Scattering Data}

\author{V.~Pandey}\email{vpandey@fnal.gov; Presented at the 23rd International Workshop on Neutrinos from Accelerators (NuFACT 2022), Salt Lake City, UT, USA, July 2022.}
\affiliation{Fermi National Accelerator Laboratory, Batavia, Illinois 60510, USA}

\begin{abstract}
A thorough understanding of neutrino-nucleus interactions physics is crucial to achieving precision goals in broader neutrino physics programs. The complexity of nuclei comprising the detectors and limited understanding of their weak response constitutes one of the biggest systematic uncertainties in neutrino experiments - both at intermediate energies affecting the short- and long-baseline neutrino programs as well as at lower energies affecting coherent scattering neutrino programs. While electron and neutrino interactions are different at the primary vertex, many underlying relevant physical processes in the nucleus are the same in both cases, and electron scattering data collected with precisely controlled kinematics, large statistics and high precision allows one to constrain nuclear properties and specific interaction processes. To this end, electron-nucleus scattering experiments provide vital complementary information to test, assess and validate different nuclear models and event generators intended to be used in neutrino experiments. In fact, for many decades, the study of electron scattering off a nucleus has been used as a tool to probe the properties of that nucleus and its electromagnetic response. While previously existing electron scattering data provide important information, new and proposed measurements are tied closely to what is required for the neutrino program in terms of expanding kinematic reach, the addition of relevant nuclei and information on the final states hadronic system.
\end{abstract}

\hfill\preprint{FERMILAB-CONF-23-015-ND}

\maketitle


\section{Introduction}

The success of current and future neutrino experiments in achieving discovery level precision will greatly depend on the precision with which the fundamental underlying process -- neutrino interaction with the target nucleus in the detector -- is known~\cite{NuSTEC:2017hzk}. To this end, electron scattering experiments have been playing a crucial role by providing high-precision data as the testbed to assess and validate different nuclear models intended to be used in neutrino experiments~\cite{Ankowski:2022thw}. 

For the accelerator-based neutrino program, such as DUNE, the primary physics goals are determining mass hierarchy and measuring precision oscillation physics including subtle effects of $\delta_{CP}$~\cite{DUNE:2020jqi}. The main challenges in constraining neutrino-nucleus scattering physics stem from the fact that neutrino energies at these experiments typically range from 100s of MeV to a few GeV where different interaction mechanisms yield comparable contributions to the cross section. One has to constrain an accurate picture of the initial state target nucleus, its response to the electroweak probe that includes several reaction mechanisms resulting into various finals state particles, and final state interactions that further modify the properties of the hadronic system created at the primary interaction vertex.

For the coherent elastic neutrino–nucleus scattering (CEvNS) program at stopped pion sources, such as at ORNL, the main source of uncertainty in evaluating the CEvNS cross section is driven by the underlying nuclear structure, embedded in the weak form factor, of the target nucleus. The recent detection of CEvNS process by the COHERENT collaboration~\cite{COHERENT:2017ipa} has opened up a slew of opportunities to probe several Beyond the Standard Model (BSM) scenarios in CEvNS experiments. In order to disentangle new physics signals from the SM expected CEvNS rate, the weak form-factor which primarily depends on the neutron density has to be known at percent level precision~\cite{Tomalak:2020zfh, VanDessel:2020epd}. 

Most of our current knowledge about the complexity of the nuclear environment - nuclear structure, dynamics, and reaction mechanisms - has been accumulated by studying electron scattering off target nuclei. The electron scattering of the nucleus, governed by quantum electrodynamics, has an advantage over the proton or pion scattering off nuclei which are dominated by strong forces. The electromagnetic interaction is well known within quantum electrodynamics and is weak compared to hadronic interaction and hence the interaction between the incident electron and the nucleus can be treated within the Born approximation, i.e. within a single-photon exchange mechanism. 

In the last few decades, a wealth of high-precision electron scattering data has been collected, over a variety of nuclei ranging from $^{3}$He to $^{208}$Pb, at several facilities including Bates, Saclay, Frascati, DESY, SLAC, NIKHEF, Jefferson Lab, etc., among others. The ability to vary electron energy, and scattering angle and hence the energy and moment transferred to the nucleus ($\omega, q$) - combined with the advancement in high-intensity electron beams, high-performance spectrometers and detectors - resulted in investigating processes ranging from quasi- elastic (QE) to the $\Delta$ resonance to complete inelastic (resonant, non-resonant, and the deep inelastic scatter- ings (DIS)) with significant details. A number of those datasets were further utilized to separate the longitudinal and transverse response functions through the Rosenbluth separation. Several decades of experimental work has provided sufficient testbed to assess and validate several theoretical approximations and predictions and hence propelled the theoretical progress staged around nuclear ground state properties, nuclear many-body theories, nuclear correlations, form factors, nucleon-nucleon interactions, etc. A web archive of accumulated data is maintained at Ref.\cite{qe_archive, Benhar_1}. 

Besides being immensely interesting in itself, electron scattering turned out to be of great importance for neutrino programs. The data collected with electron-nucleus scattering has provided the benchmark to test the nuclear models that can be further extended to neutrino-nucleus scattering. The extension of the formalism from electron-nucleus scattering, where only vector current contributes, to neutrinos require the addition of axial current contribution. Despite the fact that (unpolarized) electron scattering provides access to only vector response, the vector current is conserved between electromagnetic and weak response through conserved vector current (CVC).

While previous and existing electron scattering experiments provide important information, new dedicated measurements whose goals tie more closely with the needs of constraining neutrino-nucleus interactions physics in neutrino programs are needed. Dedicated electron scattering experiments with targets and kinematics of interest to neutrino experiments (CEvNS, supernova, and accelerator-based) will be vital in the development of neutrino-nucleus scattering physics modeling that underpin neutrino programs~\cite{Ankowski:2022thw}.

The rest of this article is structured as follows. In Sec.~\ref{sec:impact}, we briefly describe the neutrino interaction challenges faced by neutrino programs in their key physics goals. We then identify connections between electron- and neutrino-nucleus scattering physics in Sec.~\ref{sec:elec-to-neutrino}. In Sec.~\ref{sec:expt_landscape}, we present a brief summary of the current and planned electron scattering experiments that are input to various neutrino programs. We summarize in Sec.~\ref{sec:summary}.


\section{The Importance of Constraining Neutrino-Nucleus Interactions Physics}
\label{sec:impact}

In accelerated-based neutrino oscillation program, neutrino-nucleus interactions constitute one of the dominant systematic uncertainties. In the event of a neutrino oscillating from $\nu_i$ to $\nu_j$ and for a given observable topology, the observed event rate at far detector is a convolution of neutrino flux at near detector ($\phi_{\nu_{i}}$), probability of oscillation from flavor $i$ to $j$, and the 
neutrino-nucleus cross section for neutrino flavor $j$ ($\sigma_{\nu_j}$), and detection efficiency at far detector ($\epsilon_{\nu_j}$)
\begin{equation}
 \mathcal{R}(\nu_i \rightarrow \nu_j) \propto \phi_{\nu_i} \otimes P(\nu_i \rightarrow \nu_j) \otimes \sigma_{\nu_j} \otimes \epsilon_{\nu_j} \label{rate}
\end{equation}
with oscillation probability, considering simple example of two neutrino flavors, given as: 
\begin{equation}
 P(\nu_i \rightarrow \nu_j) \simeq \sin^{2}2\theta \sin^{2}\left(\frac{\Delta m^{2} L}{4E_{\nu}}\right),
\end{equation}
with $\theta$ and $\Delta m^{2}$ are mixing angle and the squared-mass difference respectively, $E_{\nu}$ is neutrino energy and $L$ is the oscillation baseline. Typically, the ratio of oscillated event rate at the far detector to the unoscillated event rate at the near detector does not cancel out flux and cross sections dependence. 

The systematic challenges in neutrino experiments are manifold. The energy of the interacting neutrino is not known, the kinematics of the interaction in the target nucleus is not known and the only known (to some degree) quantity - the topology of the final state particles and their energy - is subjected to detector type, detection thresholds, and the accuracy of particle identification and background reduction processes. The necessity of an accurate neutrino interaction recipe is essential at almost every step of the analysis. The accuracy of the measurement of the (energy-dependent) neutrino oscillation probability relies strongly on the accuracy with which a Monte-Carlo event generator can describe all neutrino-nucleus interaction types that can produce the observed event topology (that depends both on the initial and final state nuclear effects). As it stands currently, a lack of reliable nuclear recipe contributes to the main source of systematic uncertainty and is considered one of the main hurdles in further increasing the obtained precision. For current long-baseline neutrino experiments, T2K and NOvA, neutrino-nucleus interactions constitute one of the largest uncertainties. In future long-baseline neutrino experiments, DUNE and HyperK, the statistics will significantly increase and neutrino interaction systematics uncertainties will be dominant. 

In the energy regime of the accelerator-based neutrino experiments, 100s of MeV to a few GeV, several mechanisms contribute to the nuclear response: from the excitation of nuclear collective states in the lowest energy part of the spectrum, up to the deep inelastic scattering at the highest energy transfers, encompassing the quasi-elastic region, corresponding to one-nucleon knockout, and the resonance region, corresponding to the excitation of nucleon resonances followed by their decay and subsequent production of pions and other mesons. There is no unified underlying theory to describe neutrino-nucleus interactions for this broad energy range. It truly is a multi-scale, multi-process, many-body non-perturbative problem subject to complex nuclear structure and dynamics that include transitions between different degrees of freedom. One needs a description of initial state target nucleus, its response to the electroweak probe that include several reaction mechanisms resulting into various finals state particles, and final state interactions that further modify the properties of the hadronic system created at the primary interaction vertex.

Similarly, for low-energy (10s of MeV) neutrinos, the uncertainties on inelastic neutrino-nucleus interaction, the detection channel for supernova neutrinos in DUNE and HyperK is large  and is often not even quantified~\cite{DUNE:2020zfm, Gardiner:2020ulp}. Although theoretical uncertainties, primarily driven by the poorly known neutron density distributions, are relatively small in CEvNS case, percent level precision might be needed to disentangle new physics signals~\cite{Tomalak:2020zfh}. 


\section{Connecting Electron- and Neutrino-nucleus Scattering Physics}
\label{sec:elec-to-neutrino}

\begin{figure}
\includegraphics[width=0.49\textwidth]{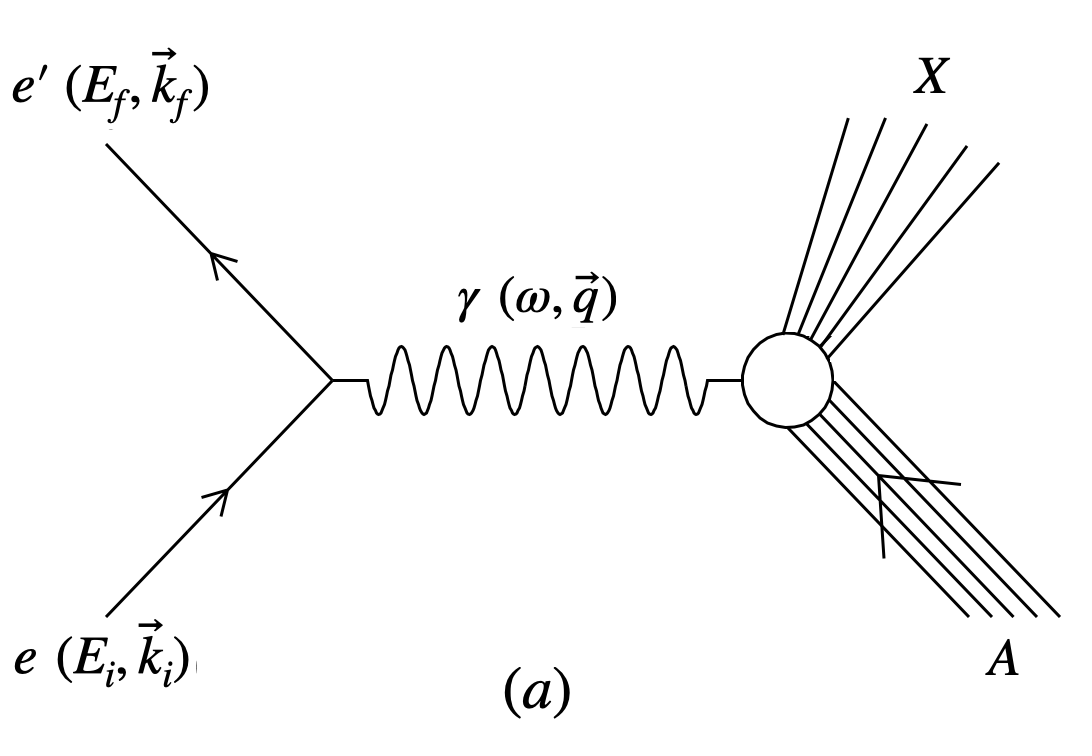}
\includegraphics[width=0.49\textwidth]{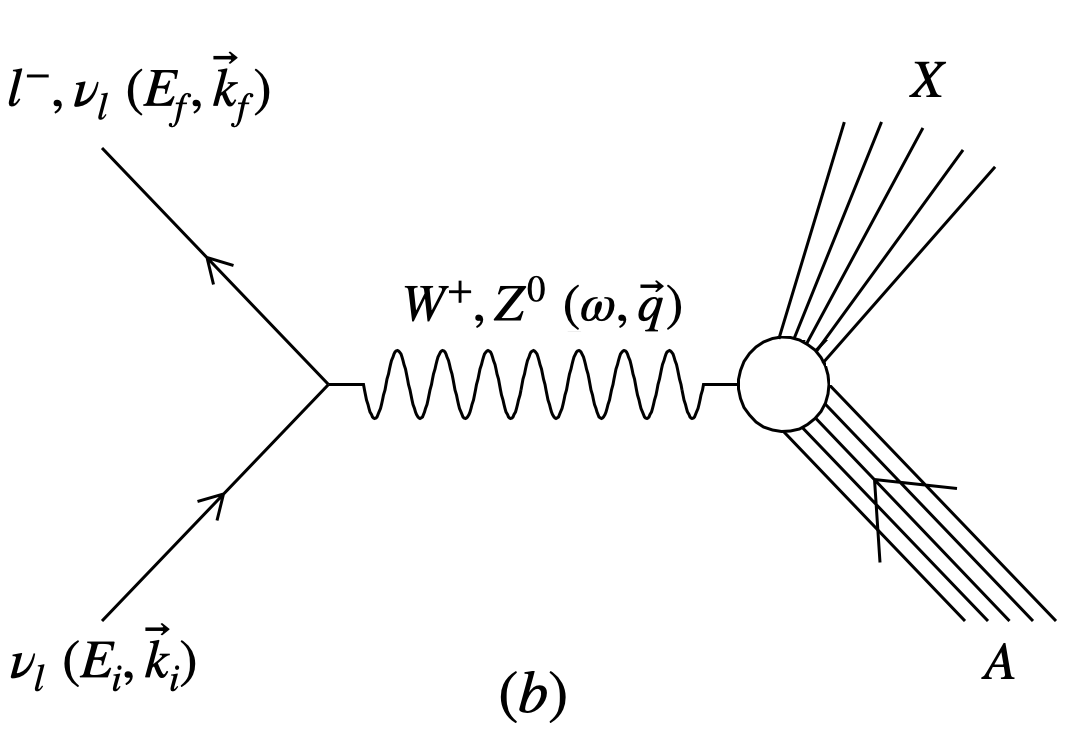}
\caption{Diagrammatic representation of (a) electron-nucleus and (b) neutrino-nucleus scattering process ($l = e, \mu, \tau$), where X represents outgoing hadronic final state.\label{fig1}}
\end{figure}   

Electron-nucleus scattering process, represented in Figure~\ref{fig1}(a), is primarily governed by electromagnetic interactions where (to first-order) the interaction is mediated by a (virtual) photon. The neutrino-nucleus scattering, represented in Figure~\ref{fig1} (b), is primarily governed by weak interaction via the exchange of a $W^{\pm}$ or $Z^0$ boson for charged and neutral
weak process, respectively. 
 
In the Born approximation~\cite{donnelly_1986} the lepton-nucleus differential cross section $d\sigma$ is proportional to the contraction of the leptonic ($L_{\mu\nu}$) and hadronic ($W^{\mu\nu}$) tensors

\begin{equation}
  d\sigma \propto L^{\mu\nu} W_{\mu\nu}
\end{equation}
with hadronic tensor written in terms of nuclear current operaters, $J$
\begin{equation}
  W_{\mu\nu} \propto \sum_f <\psi_i |J^{\dagger}_{\mu}(q)| \psi_f> <\psi_f |J_{\nu}(q)| \psi_i> \delta(E_0+\omega-E_f)
\end{equation}
where $\psi_i$ and $\psi_f$ are initial and final state wave functions, and $\omega$ and $q$ are energy and momentum transferred to the nucleus. 

Contracting the leptonic and hadronic tensor, we obtain a sum involving projections of the current matrix elements. It is convenient to choose these to be transverse and longitudinal with respect to the virtual boson direction. The electron-nucleus scattering cross section becomes
\begin{equation}
  d\sigma_e \propto V_L R_L + V_T R_T 
\end{equation}
and the neutrino-nucleus scattering cross section becomes
\begin{equation}
  d\sigma_{\nu} \propto V_{C} R_{C} +V_{L} R_{L} + 2 V_{CL} R_{CL}  +V_{T}R_{T} \pm V_{T'}R_{T'} \label{eq:neutrino}
\end{equation}
where $R$ are nuclear responses that are functions of $\omega$ and $q$. The subscripts C, L, and T correspond to coulomb, longitudinal and transverse components. The last term in Eq.~\ref{eq:neutrino}, the transverse interference term, is positive for neutrino scattering and negative for antineutrino scattering. 

The underlying nuclear physics, probed by electrons and neutrinos, is intimately connected to each other. The initial nucleus description is the same. The weak current carried by neutrinos has a vector and an axial component, while the electromagnetic current carried by electrons is purely vector. Though, the vector current is conserved (CVC) between electromagnetic and weak interactions leaving the axial nuclear response unique to neutrinos (or polarized electrons). The final state interactions effects are the same. Therefore,  various aspects of nuclear structure and dynamics influencing the neutrino-nucleus cross section can be studied in electron scattering. Any model or event generator that does not work for electron scattering would likely not work for neutrino scattering. In typical electron scattering experiments the incident beam energy is known with good accuracy, hence the transferred energy, $\omega$, and momentum, $q$, can be precisely determined by measuring the outgoing lepton kinematics. The high-precision, high statistics electron scattering data collected with precisely controlled kinematics allows to separate different processes. 

The tens-of-MeV neutrinos, from stopped pion sources or from core-collapse supernova, primarily interact via two processes: coherent elastic neutrino-nucleus scattering (CEvNS), and inelastic charged and neutral current scattering. Precise determination of Standard Model CEvNS cross section will enable new physics searches in CEvNS experiments, while precise inelastic cross section determination will enable detection of supernova signals in DUNE experiment. 

The CEvNS cross section (at tree level) is given as
\begin{eqnarray}
\label{eq_cs}
\frac{d\sigma}{dT}(E_\nu,T) &\simeq& \frac{G_F^2}{4\pi} M \left[ 1- \frac{MT}{2E_\nu^2} \right] Q^2_W F^2_W(q^2) \,,
\end{eqnarray}
where $G_F$ is the Fermi constant, $M$ the mass of the nucleus, $E_\nu$
and $T$ the energy of the neutrino and  the nuclear recoil energy, respectively. The weak form factor $F_W(q^2)$ is given as
\begin{eqnarray}
\label{eq_w}
 F_W(q^2) &=& \frac{1}{Q_W}[ NF_n(q^2) -(1-4\sin^2\theta_W)ZF_p(q^2) ] 
\end{eqnarray}
where $\theta_W$ is the Weinberg mixing angle.
In Eq.~(\ref{eq_w}) the quantities
 $F_{p}(q^2)$ and $F_{n}(q^2)$ are the proton ($p$) and neutron ($n$) form factors,  respectively.
 While the proton distributions are relatively well known through elastic electron scattering experiments~\cite{DeVries:1987atn}, neutron distributions are much more difficult to constrain. Since  $1-4\sin^2\theta_W(0) \approx 0$, the weak
 form factor becomes $F_W(q^2)\approx F_n(q^2)$. In order to disentangle new physics signals from the SM expected CEvNS rate, the weak form factor, which primarily depends on the neutron density, has to be known at percent level precision.

Recent advancements in Parity Violating Electron Scattering (PVES) experiments, utilizing polarized electron beams, provide relatively model-independent ways of determining weak form factors that can be used as direct input in determining CEvNS cross section. Both processes are described in first-order perturbation theory via the exchange of an electroweak gauge boson between a lepton and a nucleus. While in CEvNS the lepton is a neutrino and a $Z^0$ boson is exchanged, in PVES the lepton is an electron, but measuring the asymmetry allows one to select the interference between the $\gamma$ and $Z^0$ exchange. As a result, both the CEvNS cross section and the PVES asymmetry depend on the weak form factor $F_W(Q^2)$, which  is mostly determined by the neutron distribution within the nucleus. The parity-violating asymmetry $A_{pv}$ for elastic electron scattering is the fractional difference in cross section for positive helicity and negative helicity electrons.  In Born approximation $A_{pv}$ is proportional to the weak form factor $F_W(q^2)$~\cite{horowitz_2001, horowitz_1998},
\begin{equation}
    A_{pv}=\frac{d\sigma/d\Omega_+-d\sigma/d\Omega_-}{d\sigma/d\Omega_++d\sigma/d\Omega_-}=\frac{G_Fq^2|Q_W|}{4\pi\alpha\sqrt{2}Z}\frac{F_W(q^2)}{F_{ch}(q^2)}\, .
    \label{eq.Apv}
\end{equation}
Here $F_{ch}(q^2)$ is the (E+M) charge form factor that is typically known from unpolarized electron scattering.  Therefore, one can extract $F_W$ from measurements of $A_{pv}$. Note that Eq. \ref{eq.Apv} must be corrected for Coulomb distortions~\cite{Horowitz:1998vv}, though these effects are absent for neutrino scattering. 

The inelastic neutrino-nucleus cross sections in this tens-of-MeV regime are quite poorly understood. There are very few existing measurements, none at better than the 10\% uncertainty level. As a result, the uncertainties on the theoretical calculations of, e.g., neutrino-argon cross sections are not well quantified at all at these energies. Because inelastic neutrino interactions have big uncertainties, in the future it will be crucial to measure inelastic electron scattering cross sections at energies below the 50 MeV mark and use those data to calibrate theoretical models for the neutrino scattering process. Overall, we expect nuclear structure effects to be definitely larger than in CEvNS and presumably at least at the 10$\%$ level. To this end, 10s of MeV electron scattering data will be vital in constraining neutrino-nucleus interaction at this energy scale. 


\section{Current and Future Electron Scattering Experiments for Neutrino Programs}
\label{sec:expt_landscape}

\begin{table}
\centering
\begin{tabular}{|c|ccc|}
\hline
\textbf{Collaborations}	& \textbf{Kinematics}	& \textbf{Targets} &  \textbf{Scattering}\\
\hline
{\bf E12-14-012 (JLab)} & $E_e$ = 2.222 GeV & Ar, Ti & ($e,e'$) \\
{(Data collected: 2017)} & $15.5^\circ\leq{\theta_e}\leq21.5^\circ$ & Al, C & $e, p$ \\
 &$-50.0^\circ\leq{\theta_p}\leq-39.0^\circ$ &  & in the final state \\
\hline
{\bf e4nu/CLAS (JLab)} & $E_e$ = 1, 2, 4, 6 GeV& H, D, He, & ($e,e'$) \\
{(Data collected: 1999, 2022)} & ${\theta_e} > 5^\circ$  & C, Ar, $^{40}$Ca, & $e, p, n, \pi,\gamma$ \\
 &  & $^{48}$Ca, Fe, Sn & in the final state \\
\hline
{\bf LDMX (SLAC)} & $E_e$ = 4.0, 8.0 GeV & & ($e,e'$) \\
  {(Planned)} & ${\theta_e} < 40^\circ$ & W, Ti, Al & $e, p, n, \pi, \gamma$ \\
 &  &  &  in the final state \\
\hline
{\bf A1 (MAMI)} & 50 MeV $\leq E_e \leq 1.5$ GeV & H, D, He & ($e,e'$) \\
 (Data collected: 2020) & $7^\circ\leq{\theta_e} \leq 160^\circ$ & C, O, Al & 2 additional \\
 (More data planned) &  & Ca, Ar, Xe & charged particles  \\
\hline
{\bf A1 (eALBA)} & $E_e$ = 500 MeV & C, CH & ($e,e'$) \\
 {(Planned)} & ~~~~~~- few GeV & Be, Ca & \\
\hline
\end{tabular}
\caption{Current and planned electron scattering experiments. For more details, please see Ref.~\cite{Ankowski:2022thw}.\label{tab1}}
\end{table}

For over five decades, electron scattering experiments at different facilities around the world have provided a wealth of information on the complexity of nuclear structure, dynamics and reaction mechanisms. Decades of experimental work has provided a vital testbed to assess and validate theoretical approximations and predictions that propelled the theoretical progress staged around. A large data set of high precision electron-nucleus scattering exist, meant to study various nuclear physics aspects, covering many nuclei and wide energy ranges corresponding to different reaction mechanisms. While previous and existing electron scattering experiments provide important information, new dedicated measurements whose goals are tied to the needs of neutrino programs are needed. New data can expand relevant kinematic reach, the addition of relevant nuclei and the information on the final states hadronic system.

In Table~\ref{tab1}, we present a summary of the current and planned electron-scattering experiments. These electron scattering experiments are primarily motivated by the needs of the accelerator neutrino experiments. They include complementary efforts that cover a broad range of kinematics and carry a varied level of particle identification and other detection capabilities. The work is mainly done with a cross-community effort of nuclear and high-energy physicists. For more information on details of individual experiments, we refer readers to a recent Snowmass white paper, Ref.~\cite{Ankowski:2022thw}. 

\begin{figure}
\includegraphics[width=1.0\textwidth]{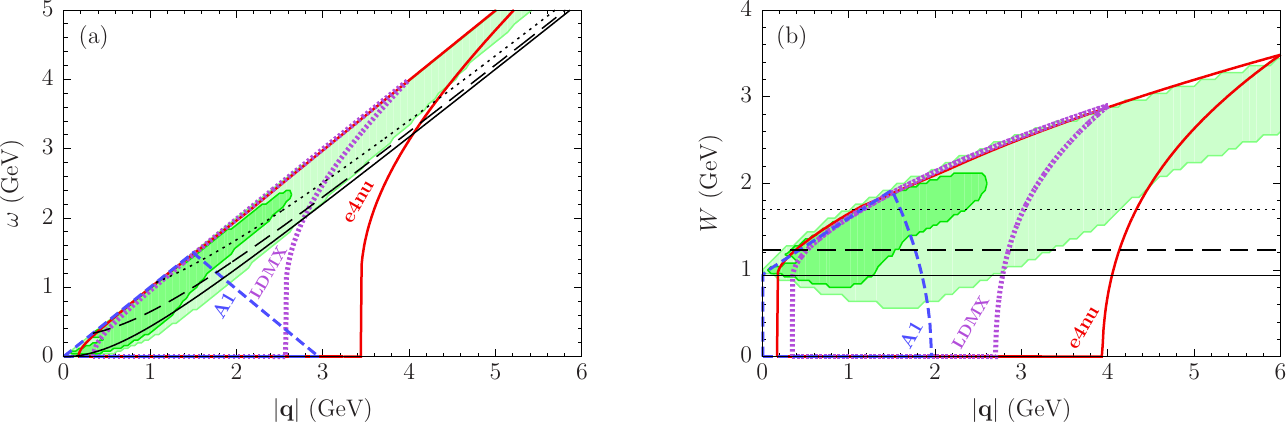}
\caption{Kinematic coverage of the ongoing and planned electron scattering experiments for electron scattering on targets including argon and titanium presented in the (a) $(|\mathbf{q}|, \omega)$ and (b) $(|\mathbf{q}|, W)$ planes. The thin solid, dashed, and dotted lines correspond to the kinematics of quasielastic scattering, $\Delta$ excitation, and the onset of deep-inelastic scattering at $W = 1.7$ GeV on free nucleons. Figure taken from Ref.~\cite{Ankowski:2022thw}.\label{fig2}}
\end{figure}   

This kinematics is then presented in Fig.~\ref{fig2}, where it is overlaid on the regions expected to contain 68\% (light shaded) and 95\% (dark shaded) of charged current interactions of muon neutrinos with argon in the DUNE near detector~\cite{DUNE:2021cuw}, as estimated using GENIE 3.0.6. The e4nu experiment at JLab~\cite{CLAS:2021neh} employs a broad range of energies and has the potential to study a significant phase space of DUNE kinematics.  The beam energy of the LDMX experiment at SLAC~\cite{Ankowski:2019mfd}, 4 GeV, is closely corresponding to the average neutrino energy in DUNE, and can perform extensive studies of pion production. A1 collaboration at MAMI covers a broad range of scattering angles---from $7^\circ$ to $160^\circ$---and beam energies---from $\sim$50 MeV to 1.5 GeV--- and would be able to perform extensive studies of the quasielastic and $\Delta$ peaks. In these experiments, in general, a lot of attention will be given to measuring  exclusive cross sections.

In Table~\ref{tab2}, we present a summary of the current and planned PVES experiments. These experiments probe complementary information for CEvNS experiment in constraining weak form factor of the nucleus. While for tens-of-MeV inelastic neutrino scattering, there is currently no ongoing program though the potential exists for a lower energy electron beam at MESA at Mainz. Dedicated electron scattering experiments with targets and kinematics of interest to low-energy neutrino experiments will be crucial in achieving the precision goals of low-energy neutrino programs. For more information, we refer readers to a recent Snowmass white paper, Ref.~\cite{Ankowski:2022thw}.

\begin{table}
\begin{tabular}{|c|cccc|}
\hline
\textbf{Collaborations}	& \textbf{Target}	& \textbf{$q^2$ (GeV$^2$)} &  \textbf{$A_{pv}$ (ppm)} &  \textbf{$\pm\delta R_n$ (\%)}\\
\hline
PREX & $^{208}$Pb & 0.00616 & $0.550\pm0.018$ & 1.3 \\
CREX & $^{48}$Ca & 0.0297 & & 0.7\\
Qweak & $^{27}$Al & 0.0236 & $2.16\pm 0.19$ & 4\\
MREX  & $^{208}$Pb & 0.0073 & & 0.52\\
\hline
\end{tabular}
\caption{Parity violating elastic electron scattering experiments. For more details, please see Ref.~\cite{Ankowski:2022thw}.\label{tab2}}
\end{table}


\section{Summary}
\label{sec:summary}

Neutrino physics has entered a precision era and exciting neutrino experimental programs at low
and high energies can lead to discoveries. The importance of constraining systematics resulted from
neutrino-nucleus interaction physics in key neutrino measurements, in particular at accelerator-based
experiments, cannot be overstated. To this end, the electron scattering experiments play a key role in 
constraining underlying nuclear physics in nuclear models and event generators intended to be used in neutrino experiments.

Electron and neutrino interactions carry many similarity in underlying relevant physical processes, and electron scattering data collected with precisely controlled kinematics, large statistics and high precision allows one to constrain nuclear properties and specific interaction processes. Electron scattering data provide the necessary testbed to assess and validate different nuclear models
intended to be used in neutrino experiments. While previously existing electron scattering data 
provide important information, new and proposed measurements whose goals are closely tied to the needs of neutrino program in terms of expanding kinematic reach, the addition of relevant nuclei and information on the final states hadronic system are vital. The NP-HEP cross-community collective efforts are playing a key role in this endeavour.


\acknowledgments{VP is grateful to the organizers of the NuFACT 2022 workshop for the invitation and hospitality. This manuscript has been authored by Fermi Research Alliance, LLC under Contract No. DE-AC02-07CH11359 with the U.S. Department of Energy, Office of Science, Office of High Energy Physics.}



\end{document}